\documentclass[11pt]{article}
\usepackage{inputenc}
\usepackage[T1]{fontenc}
\usepackage{lmodern}
\usepackage[ngerman,UKenglish]{babel}
\usepackage[a4paper,height=24.5cm]{geometry}
\DeclareUnicodeCharacter{012D}{\u{\i}}
\DeclareUnicodeCharacter{012D}{\u\i}
\usepackage{graphicx}%
\usepackage{amsmath,amssymb,amsfonts,soul}
\usepackage[twoside]{fancyhdr}%
\usepackage{mathrsfs}%
\usepackage[title]{appendix}%
\usepackage{xcolor}%
\usepackage{textcomp}%
\usepackage{xspace}
\usepackage[font=small]{caption}
\usepackage[numbers]{natbib}
\usepackage[hyphens]{url}
\usepackage{doi}
\usepackage{hyperref}

\newcommand{\ie}{i.\,e.\@\xspace}
\newcommand{\eg}{e.\,g.\@\xspace}

\begin{document}

\title%
{\fontsize{15}{20}\selectfont\bf The EPR Paradox of Quantum Mechanics in the Light of Four Unpublished Letters between\\ A.~Einstein and the Mathematician \nolinebreak J.\,L.\,B.~Cooper}
\newcommand{\shorttitle}{The EPR Paradox of Quantum Mechanics}
\author{P.\,L.~Butzer\thanks{Lehrstuhl A für Mathematik, RWTH Aachen, 52056 Aachen, Germany,\hfil\rule{0pt}{0pt} email:~\url{butzer@rwth-aachen.de}}
\and\quad D.\,E.~Edmunds\thanks{Department of Mathematics, University of Sussex, Brighton, BN1~9QH, U.\,K., \hfil\rule{0pt}{0pt} email:~\url{davideedmunds@aol.com}}
\and\quad G.~Roepstorff\thanks{email:~\url{roep@arcor.de}}
\and\quad G.~Schmeisser\thanks{Department of Mathematics, University of Erlangen-Nuremberg, 91058~Erlangen, Germany,\hfil\rule{0pt}{0pt} email:~\url{schmeisser@mi.uni-erlangen.de}}
\and\quad R.\,L.~Stens\thanks{Lehrstuhl A für Mathematik, RWTH Aachen, 52056 Aachen, Germany,\hfil\rule{0pt}{0pt} email:~\url{stens@matha.rwth-aachen.de}}}
\newcommand{\authors}{P.\,L.~Butzer, D.\,E.~Edmunds, G.~Roepstorff, G.~Schmeisser, R.\,L.~Stens}
\date{}                              
\maketitle

\setlength{\headheight}{15pt}
\pagestyle{fancy}    
\fancyhf{}                           
\fancyfoot{}
\fancyhead[EL,OR]{\thepage}
\fancyhead[RE]{\authors}
\fancyhead[lO]{\shorttitle}

\section*{\fontsize{11}{11}\selectfont Abstract}
This paper presents correspondence between Albert Einstein and the mathematical analyst J.\,L.\,B.~Cooper on the Einstein-Podolsky-Rosen (EPR) paradox of quantum theory published in 1935. Two letters written by Cooper, and the replies from Einstein, all written between October and December 1949, are retyped from the original ones. Furthermore, Einstein's second letter, which he wrote in German, is translated into English. The lack of agreement, arising from very different points of view, is analysed, taking into account the complex underlying mathematical and physical factors that arise naturally in connection with the EPR paradox.

\section*{\fontsize{11}{11}\selectfont Keywords}Quantum Mechanics, Einstein-Podolsky-Rosen paradox, history of physics and mathematics


\section{Introduction}\label{sec_introdution}

The Einstein--Podolsky--Rosen paradox of quantum theory (EPR paradox, for short) has often been considered and discussed in the literature; for recent papers see, e.\,g., \cite{Blaylock_2010,Fine_2016,Goette_et_al_2016,Handsteiner_et_al_2017,Kupczynski_2016,Levi_1015,Peise_et_al_2015}. Here we present correspondence between the mathematician Lionel Cooper and Albert Einstein on the EPR paradox, illuminate its historical background and comment on the Einstein--Cooper letters.

\begin{figure}
 \includegraphics[width=\textwidth]{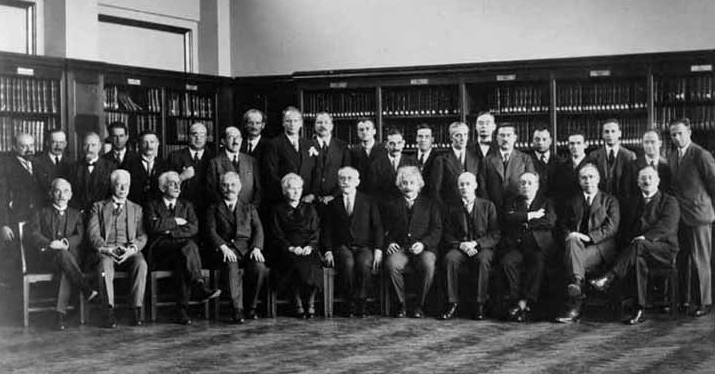}
  \caption{Sixth Solvay Conference on Physics, Brussels, 1930. Standing, from left to right E.~Herzen, É.~Henriot, J.~Verschaffelt, C.~Manneback, A.~Cotton, J.~Errera, O.~Stern, A.~Piccard, W.~Gerlach, C.~Darwin, P.\,A.\,M.~Dirac, E.~Bauer, P.~Kapitsa, L.~Brillouin, H.\,A.~Kramers, P.~Debye, W.~Pauli, J.~Dorfman, J.\,H. Van~Vleck, E.~Fermi, W.~Heisenberg; sitting, from left to right Th.~De~Donder, P.~Zeeman, P.~Weiss, A.~Sommerfeld, M.~Curie, P.~Langevin, A.~Einstein, O.~Richardson, B.~Cabrera, N.~Bohr, W.\,J.~De~Haas}\label{fig_Solvay}
 \end{figure}

The EPR article \cite{EPR} originated from Einstein's reflections on a ``simple thought-experiment'' he proposed at the Sixth Solvay Congress, held in Brussels, October 20--25, 1930, and presided over by Paul Langevin. Although it was devoted to Magnetism, a major subject of discussion was the foundations of quantum mechanics. Thirty two of the world's leading physicists participated; see Figure~\ref{fig_Solvay}.
According to Einstein's letter to Paul Epstein of November~10, 1945 \cite[p.~234]{Jammer_1974}, that experiment concerned an ideally reflecting photon box which contains a clock which appears able to time precisely the release in the box of a photon with determinate energy. If this were feasible, it would appear to challenge the unrestricted validity of the Heisenberg uncertainty principle that sets a lower bound on the simultaneous uncertainty of energy and time. This was a central component in the unrealistic interpretation of the wave function.

Max Jammer \cite[pp.~166--181]{Jammer_1974} (see also Don~Howard \cite{Howard_1990} and Arthur Fine \cite{Fine_2016}) describes how Einstein's thoughts about this experiment, and Bohr's objections to it, evolved into a different photon-in-a-box experiment, one that allows an observer to determine either the momentum or the position of the photon indirectly, while remaining outside, sitting on the box. Jammer associates this with the distant determination of either momentum or position, which is the core of the EPR article.

Now we want to introduce Cooper to the reader. A detailed biography can be found in \cite{Butzer-Schmeisser-Stens_2016}; see also \cite{Butzer_1981,Edmunds_1981},  \url{https://mathshistory.st-andrews.ac.uk/Biographies/Cooper/}, and \url{https://en.wikipedia.org/wiki/Lionel Cooper (mathematician)}.

Jacob Lionel Bakst Cooper, born on December 27, 1915 in Beaufort West, South Africa, was a distinguished analyst who made notable contributions to functional analysis and Fourier theory; in 1949 he was awarded the Berwick Prize of the London Mathematical Society. South African by birth, he came to Oxford as a Rhodes scholar and obtained a D.Phil.\ under the supervision of the celebrated mathematician E.\,C.\ Titchmarsh. Posts at Birkbeck College London, Cardiff, Caltech, Toronto and Chelsea College, London followed; he died on August 8, 1979.

After his death, his correspondence was found to include two letters from Einstein, dated 31 October and 18 December, 1949, dealing with the EPR paradox. No trace of communications from Cooper to Einstein could be found, but an internet search showed that the Einstein Archive in Jerusalem had two letters from Cooper, dated 11 October and 19 November, 1949, that had plainly stimulated Einstein's responses; they had not been digitalised. The Archive was keen to obtain copies of Einstein's letters, which they did not have, and offered Cooper's letters in return. There was no further exchange of views between Einstein and Cooper. We feel that the complete correspondence thus obtained is of sufficient interest to share with the scientific community and give it below.

The paper is organized as follows.
In Section~\ref{sec_historical} we take a  historical view of the EPR paradox by giving an account of the reactions by some of Einstein's eminent contemporaries.
In Section~\ref{sec_retyped} we present the four letters constituting the correspondence between Cooper and Einstein, retyped from the original ones. Although Einstein in 1949 had already lived for 16 years in USA, it seems that he still felt more at ease with  German since in his second letter he replied Cooper in that language. For the reader's convenience we add an English translation. After that letter the correspondence ceased. Einstein and Cooper did not come to an agreement. The reason may have been the different views of a physicist and a mathematician. Therefore, in Section~\ref{sec_comments}, where we give comments on the correspondence, we do this separately from a physicist's and a mathematician's point of view.

\section{Historical background---Reactions on the EPR paper}\label{sec_historical}

Einstein, who came from Berlin to the Institute for Advanced Study at Princeton in the fall of 1933 on the invitation of Abraham Flexner, its founder and first director, sought young mathematicians or physicists to assist him with his work, especially after Walter Mayer, who came with him, obtained an independent position. He picked on Boris Podolsky and Nathan Rosen, and already in the May~15, 1935 issue of \emph{Physical Review} the three co-authored their famous \emph{Can Quantum-Mechanical Description of Physical Reality be Considered Complete?} \cite{EPR}. It is well-known that this EPR paper, as it came to be called, was actually written by Podolsky ``after many discussions'' between the authors but Einstein, who apparently did not see the final draft, right after publishment complained that it was ``smothered by formalism (Gelehrsamkeit)''. Einstein himself published four versions of the EPR incompleteness argument, the first, \emph{Physik und Realität,} already in 1936 \cite{Einstein_1936a_deutsch} (transated into English by J.~Picard \cite{Einstein_1936a_english}), where it is explicitly referred to as a ``paradox''. The others appeared in 1949, drafted in 1947, \cite{Einstein_auto}; another in his \emph{Reply to criticisms} in the same volume \cite{Einstein_reply}, and the fourth in a special issue of \emph{Dialectrica} of 1949 \cite{Einstein_1948_Dialectica}. Further versions are found in his correspondence, in particular that with Erwin Schrödinger%
\footnote{Erwin Schrödinger (1887--1961), born in Vienna, studied 1906--10 at the University of Vienna, being under the influence of Fritz Hasenöhrl, Franz Exner and K.\,W.\,F.\ Kohlrausch. After academic positions in Jena, Stuttgart and Breslau, he spent six years in Zurich (replacing von Laue), and in 1927 became Planck's successor at Berlin. With Hitler's coming to power he resigned and accepted a fellowship at Oxford. After spending 1936--38 again in Austria, in Graz, he moved to the new Institute for Advanced Studies at Dublin, where he retired in 1955.}
of June 1935, and Lionel Cooper of October and December 1949; see also \cite{Sauer_2007} in this respect.

The EPR text consists of four parts (see Jammer \cite[p.~181/182]{Jammer_1974}): an epistemo\-logical-metaphysical preamble; a general characterization of a quantum mechanical description in terms of wave functions; the application of this description to a specific example; and fourthly, a conclusion drawn from parts one and three. In part two it is pointed out that of two physical quantities which are represented by noncommuting operators the precise knowledge of one of them precludes such a knowledge of the other. This part is essentially concerned with the logical disjunction of two assertions or options, that either

\medskip
\noindent
i) the quantum mechanical description of reality given by the wave function is incomplete;

\medskip
\noindent
or

\medskip
\noindent
ii) that incompatible quantities, the operators of which do not commute---like a coordinate of position and linear momentum in that direction---cannot have simultaneous reality.

\medskip
It follows that if option i) failed, thus if quantum mechanics were complete, then incompatible quantities cannot have
real values simultaneously.

A first premise of the three authors, later to be justified, is that either i) or ii) must hold. Their second premise, also to be justified, is that if quantum mechanics were complete, then incompatible quantities could have simultaneous real values. One can well understand Einstein's disappointment with this highly abstract argument. The three authors always regarded their arguments as conclusive evidence for the incompleteness of the quantum mechanical description of physical reality.

\bigskip

Concerning the subsequent discussion of the paradox, Lionel, in \cite{Cooper_1950b}, cited Bohr \cite{Bohr_1935_Phys-Rev}, Furry%
\footnote{Wendell Hinkle Furry (1907--1984), born in Prairieton, Indiana, who received his PhD under James H.~Bartlett from the University of Illinois in 1932, was an instructor in Kemble's department at Harvard from 1934 to 1937. He continued teaching at Harvard, served as chairman of physics, worked on radar propagation at MIT's Radiation Laboratory during wartime, and retired as emeritus in 1977. He edited and translated Russian physics journals and assisted I.~Emin in writing up his \emph{Russian-English Physics Dictionary} (Wiley, 1963).}
\cite{Furry_1936_5,Furry_1936_6}, Schrödinger \cite{Schroedinger_1935_Cambridge, Schroedinger_1935_Natur}, as well as the texts of Kemble%
\footnote{Edwin Crawford Kemble (1889 --1984), born in Delaware, Ohio, who obtained his PhD at Havard under Percy Bridgman (and G.\,W.\ Peirce) in 1917, spent his whole career since 1919 at Havard, apart from 1927--28 as a Guggenheim Fellow in Munich with Sommerfeld, and Göttingen with Born. His students included J.~van Vleck, R.\,S.\ Mulliken, Slater, J.\,R.\ Oppenheimer, L.~Dunham. During WW~II he worked on acoustic detection of submarines and the investigation of German nuclear energy efforts (Operation Alsos).}
\cite{Kemble_1937} and Reichenbach \cite{Reichenbach_1944}. The first reaction to appear in a journal was that by Edwin C.~Kemble \cite{Kemble_1935} who, while completing the manuscript of his book \cite{Kemble_1937}, expressed the opinion that the EPR argument is not sound and that the problem raised by its authors was a verbal question, completely irrelevant to the work of a physicist whose job is ``to describe the experimental facts in his domain as accurately and simply as possible, \dots''.

Some months after the publication of the EPR paper, Bohr published his response \cite{Bohr_1935_Phys-Rev}. Although he accepted much of EPR without argument, the conclusion he drew from the proposal was opposed to those of Einstein. Bohr emphasizes that the question is to what extent we can trace the interaction between the particle being measured and the measuring instrument. His main focus of attack was the Criterion of Reality which, he says, contains an ambiguity in the phrase ``without in any way disturbing a system'', an assumption which was key to the conclusion drawn by Einstein. Although Bohr concedes that the measured system in EPR would not create any ``mechanical disturbance'' on the other system, he considers that there still remains ``an influence on the very conditions which define the possible types of predictions regarding the future behavior of the system''.

The next to contribute  was Schrödinger, who  became interested in the EPR paper during his Oxford fellowship. His paper \cite{Schroedinger_1935_Cambridge}, submitted to the \emph{Cambridge Phil.\ Society} and communicated  on October 1935 by  Max Born, then at Cambridge, can in many aspects be regarded as the very antithesis of Bohr's article. According to Jammer \cite[pp.~212--221]{Jammer_1974}, Schrödinger agreed with the assumptions underlying the view of EPR, but instead of satisfying himself with the epistemological conclusions drawn by the authors, he pursued the mathematical aspects of their examples and showed in his elaborations that the conceptual situation is even more complicated than envisaged by Einstein et al. In fact, for Schrödinger, it was not only a matter of incompleteness of the theory but a manifestation of a serious flaw in its very foundations.The climax of his contributions to the paradox was his second paper to the Cambridge Society, communicated by Dirac on October 1936 \cite{Schroedinger_1936_Cambridge}.

Independently of Schrödinger's work, Furry published two papers in  the \emph{Physical Review} of 1936 \cite{Furry_1936_5,Furry_1936_6}, the latter of which was discussed in great deal by Lionel \cite[p.~624]{Cooper_1950b}. He writes: ``The interesting point about [Furry's] arrangement is that electrons [projected along the $x$-axis with total momentum in the $x$-direction] have to be separated, after collision, by a barrier perpendicular to the $z$-axis, and that it is therefore the $z$-component of momentum which is no longer self-adjoint after measurement, not the $x$-component.'' Furry's results are almost identical to those of Schrödinger from the mathematical point of view, but with an exactly diametrical interpretation to that of Schrödinger. For Furry, the decisive issue in the difference of opinion between Einstein et al.\ and Bohr is that the three took it for granted that a system has independent real properties and is in a definite quantum mechanical state as soon as it is free from mechanical interference. In Furry's view the irreconcilability of ordinary quantum formalism with such an assumption is the content of the EPR paradox. Furry does not attempt to refute the EPR argument but suggests a specific experiment by means of which a basic assumption underlying the approach of Einstein et al.\ can be tested.  An experiment, compatible with Furry's hypothesis, was carried out by Chien-Shiung Wu and Irving Shakov at Columbia University in 1949.

There were several further reactions to the  EPR paradox: we mention them briefly. Hugh C.~Wolfe of the City College of New York proposed in 1936 a simple solution  of the paradox, similar to an idea frequently  expressed by Heisenberg. Henry Margenau (1901--1997) of Yale University suggested in 1935/36 that several conceptual difficulties in the quantum mechanical description can be eliminated by rejecting the projection postulate, and sent the preprint to Einstein, who also replied to him. His second paper of 1937  led  Margenau to a statistical interpretation of quantum mechanics, not much different from that of Popper\footnote{Sir Karl Raimund Popper (1902--1994), born in Vienna, who earned a doctorate in psychology under Karl Bühler 1928, taught secondary school from 1930 to 1936. By 1934 he had already published his first book, \emph{Logik der Forschung}. The rise of Nazism led Popper to emigrate 1937 to Christchurch, New Zealand. Here he wrote his influential \emph{The Open Society and its Enemies}. He moved to England in 1949, finally to become 1949 Professor of Logic and Scientific Method at the University of London, retiring in 1969. He is regarded as one of the greatest philosophers of science of the 20th century and is admired for the scope of his intellectual influence. He won many awards and honours, including the Knighthood by Queen Elizabeth in 1965, the Lippincott Award, the Sonning Prize, and Fellowships in The Royal Society, The British Academy, The London School of Economics, Cambridge University and Charles University, Prague.}.

After WW II Paul Epstein decided to study the problem of physics and reality, a credo brought up by Einstein in
1936.

\section{The Cooper-Einstein correspondence (retyped)}\label{sec_retyped}
{\parindent0pt \parskip.4\baselineskip
\subsection*{Cooper's first letter of October 11, 1949}
%
Dear Professor Einstein,

I would like to write to you concerning some ideas I have had in connection with the wellknown paradox formulated by yourself, Podolsky and Rosen in ``Physical Review'' for 1935. I hope you will be sufficiently interested in the subject to pardon my presumption in writing to you.

As you will recall, the paradox depends considering the information about a system I got by measurements on a system II which formerly interacted with I but is now separated from it: the paradox being that we can apparently arrange to measure both position or momentum on I by suitable measurements on II and then on I, or, alternatively that we can control the state of I by the measurements we choose to do on II in spite of their separation.

It seems to me that the paradox can be removed by the following arguments. If two quantum mechanical systems are to be genuinely separated, there must be an impassable barrier between them, an infinity potential wall, by which each is confined to a half-space, for if there is not such an infinite barrier. there is a finite probability of interaction and hence the wave functions cannot be separated. Now if the infinite barrier is say in the plane $x=0$, then there is no self-adjoint operator to represent the momentum in the $x$-direction: the operator $h/2\pi i\ \partial/\partial x$ which represents momentum in a doubly infinite line is hermitian and maximal, but not self-adjoint (or hypermaximal in von Neumann's terminology.) Consequently the theorem concerning the resolution of the identity -- Dirac's ``Representation Theorem'' on the representation of every state as a sum of eigenstates -- breaks down, and with it the normal theory of measurement for the operator. The arguments of your paper are based on this theorem, and so cannot be carried through; the same applies to the similar arguments of Schrödinger's papers on the subject.

The fact that the momentum in a half-space has not a self-adjoint operator to represent it is odd enough, but is wellknown and is not a new paradox. The same difficulty arises for the radial momentum.

The whole thing turns on the distinction between self-adjoint and more general hermitian operators, and I have been trying to find some physically significant characterisation of the distinction. The best I have got is as follows:

Let us understand by a ``history'' some linear law which assigns the state of a system at an initial instant states corresponding to all values of time in some interval -- \ie to $\varphi$ a function $\psi(t)= U_t\varphi$ of time for $t$ in some interval, $U_t$ being a linear operator.

We call an operator $R$ a constant of history if its expectation value $\big(R\psi(t),\psi(t)\big)$is always constant.

Then a hermitian operator $H$ is selfadjoint if and only if there exists a history defined for all $t$, $-\infty<t<\infty$, so that the operators constant in the history are those and only those which commute with $H$.

On the other hand, $H$ is maximal, but not selfadjoint, if such a history can be defined for all $t\ge 0$, but not for $t<0$, or can be defined for all $t\le 0$, but not $t>0$.

The application to the momentum in a half-line is fairly obvious: if a particle has constant momentum, it must either be moving to the barrier, so that it can have only a finite future, or must be coming from it, so can have only a finite past, with the momentum constant.

I expect that I shall publish these ideas in due course: but in view of the time publication takes here, I thought you might be interested in this brief statement. Needless to say, I should be extremely grateful if you could let me know of any comments or criticisms, if you find time.

With very sincerest respects,

Yours truly,\\
L.~Cooper.

\subsection*{Einstein's reply of October 31, 1949}

My dear Mr.~Cooper:

Thank you for you letter of October 11t. It is not my opinion that there is a logical inconsistency in the quantum theory itself and the ``paradoxon'' does not try to show it. The intention is to show that statistical quantum theory is not compatible with certain principles the convincing power of which is independent of the present quantum theory.

These is the question: Does it make sense to say that two parts $A$ nad $B$ of a system do exist independently of each other if they are (in ordinary language) located in different parts of space at a certain time, if there are no considerable interactions between these parts (expressed in terms of potential energy) at the considered time?

Thereby it is not supposed that there is a wall of separation potential energy between $A$ and $B$. I mean by ``independent of each other'' that an action on $A$ has no immediate influence on the part $B$, In this sense I express a principle a)

a) independent existence of the spacially separated.

This has to be considered with the other thesis b)

b) the $\psi$-function is the complete description of an individual physical situation.

My thesis is that a) and b) cannot be true together, for if they would hold together the special kind of measurement concerning $A$ could not influence the resulting $\psi$-function for $B$, (after measurement of a).

The majority of quantum theorists discard a) tacitly to be able to conserve b). I, however, have strong confidence in a) so that i feel compelled to relinquish b).

By same mail I am sending you a short paper expressing my view on this question,

Sincerely yours,\\
Albert Einstein.

\subsection*{Cooper's second letter of November 19, 1949}

Dear Professor Einstein,

Thank you for your letter of the 31st October, and the offprints of your Dialectica paper, which was great interest.

I think that I did not make it clear why it seems necessary to consider the case auf electrons separated by an infinite potential barrier: if you will permit me I  will explain this here.

The the point is that in the nonrelativistic form of quantum mechanics -- which is the only form sufficiently developed for your argument to be presented at all -- there is no way of separating particles save by an infinite potential barrier. Suppose that at an initial instant it is known definitely that two particles are on intervals of the $x$-axis which are as far apart as you like. Then, since the particles are known to be confined each to final space of the line, the wave functions and terms of $x$ vanish for each particle outside a finite interval, and the momentum wave function, being the Fourier transform of the $x$-wave function, is an analytic function, and so can have only isolated zeroes. The is therefore a finite probability that the momentum will have values in any interval, however large; consequently there is a finite probability that the particles will be together, and hence interacting , at any instant after the initial instant, no matter how far apart they are initially or how short the time interval allowed for them to come into interaction.

Thus the measurements performed on part $A$ of a system, separated from $B$ by a considerable distance, effect our knowledge of coordinate and momentum for $B$ not by electromagnetic forces -- but because of the fact that an uncertainty arises as to whether $B$ has been in collision with $A$, come near enough to be affected by its field.

It is evident as long as $A$ and $B$ are not separated, any set of measurement on the two 0f them will give results for position and momentum of $B$ whose uncertainties have the product $h/2\pi$. This quantity does nor depend on the field of forces between $A$ and $B$; nor on their distance apart, and it must have the same value even  when $A$ and $B$ are far apart so that the forces acting from one to the other are negligible. From this it follows that it is the uncertainty about the possibility of $A$ and $B$ moving together, and not the direct field of force, which causes this uncertainty $??? p+x$ for $B$.

The argument about separated systems can therefore, logically, be applied only when there is a barrier which makes it impossible or the two to move together; or so it seems to me. In that case the difficulty which I stated in my last letter, that the momentum operator has no representation theorem, arises.

With many thanks,

Yours sincerely,\\
L.~Cooper.

\subsection*{Einstein's reply in German of December 18, 1949}
Dear Mr.\ Cooper:

Please, excuse me that I did not find the time to answer your letter of November 19th in English.
\begin{selectlanguage}{ngerman}

Ich habe Ihren ersten Brief völlig verstanden. Der Fall der Existenz einer isolierenden Barriere von potentieller Energie ist aber von keinem Interesse für die Frage, die ich in dem Artikel der \glqq Dielektrica\grqq\ gestellt habe. Es soll nur solche potentielle Energie vorhanden sein, welche eine zeitweise Wechselwirkung der beiden Korpuskeln darzustellen gestattet.

Ich meine auch nicht, dass die Quantentheorie an einem \underline{inneren} Widerspruch leidet. Auf was es ankommt ist dieses: Wie steht die Quantentheorie zu dem Grundsatz, dass es keine unmittelbare momentane Fernwirkung zwischen zwei Raumteilen gibt. Meine Behauptung ist, dass die Quantentheorie an diesem Prinzip nicht festhalten kann, wenn sie daran festhält, dass die $\psi$-Funktion als eine im Prinzip \underline{vollständige} Darstellung einer physikalischen Situation aufzufassen ist.

Nun folgende Bemerkungen:

1) Wenn man den Fall betrachtet, dass $\psi(x_1,x_2)$ (zu einer gewissen Zeit) so beschaffen ist, dass $\psi=0$, wenn nicht $x_1$ in einem gewissen Intervall und $x_2$ in einem \underline{anderen} Intervall ist, so ist dies als Grenzfall aufzufassen. Der analytische Character der $\psi$-Funktion erlaubt natürlich nicht, dass $\psi$ \underline{genau} verschwindet für irgend ein endliches $(x_1,x_2)$ Gebiet. Aber dies kam \underline{mit beliebiger Näherung} realisiert gedacht werden; dies genügt für unsere Überlegung. Dies bedeutet, dass wir sicher wissen, dass die beiden System-Teile räumlich voneinander getrennt sind.

2) Eine zusätzliche (vollständige) Messung am ersten Teilsystem liefert die $\psi$-Funktion des zweiten Teilsystems. Die \underline{Art} der Messung bestimmt die Art der $\psi$-Funktion des zweiten Teilsystems. Verschiedenartige Messungen am ersten Teilsystem ergeben \ul{verschieden\-artige} $\psi$-Funktionen für das zweite Teilsystem.

3) Verschiedene $\psi$-Funktionen beschreiben verschiedene reale Zustände. (Voraussetzung: dass die $\psi$-Funktion ein System vollständig beschreibt). Also hängt der reale Zustand des zweiten Teilsystems davon ab, was für eine Messung ich am ersten Teilsystem vornehme. Dies ist gleichbedeutend mit der Annahme einer unvermittelten Fernwirkung. Dies sollte gezeigt werden.

4) Es handelt sich also nicht um einen inneren Widerspruch der Quantentheorie, sondern um die Behauptung, dass folgende Alternative besteht.

Entweder ist die Beschreibung durch die Quantentheorie als eine unvollständige Beschreibung (des individuellen Systems) anzusehen

Oder man muss eine unvermittelte Fernwirkung annehmen

(Oder man kann auch beides annehmen).

Sincerely yours,\\
Albert Einstein.

\end{selectlanguage}

\subsection*{The English translation of Einstein's second letter}


Dear Mr. Cooper:

Please, excuse me that I did not find the time to answer your letter of November 19th in English.

I fully understood your first letter. However, the case  of the existence of an isolating barrier of potential energy is of no interest for the question I posed in the article in ``Dialectica''. Only such potential energy should be present which permits a temporary interaction of the two particles.

I also do not think that quantum theory suffers from an \underline{internal} contradiction. What matters is this: How does quantum theory get along with the principle that there does not exist an immediate momentary action-at-distance between two spatial parts. My assertion is that quantum theory cannot adhere to this principle if it retains the opinion that the $\psi$-function has to be understood as an essentially \underline{complete} representation of a physical situation.

Now the following remarks:

1) If one considers the case that $\psi(x_1, x_2)$ (at a certain time) is such that $\psi=0$ unless $x_1$ is in a certain interval and $x_2$ in some \underline{other} interval, then this has to be understood as a limiting case. The analytic character of the $\psi$-function does not, of course, any permit that $\psi$ vanishes \underline{exactly} on some finite $(x_1, x_2)$-region. But this can be thought to be realized \underline{with any approximation}; this is sufficient for our consideration. This means that we know for sure that the two parts of the system are spatially separated.

2) An additional (complete) measurement on the first subsystem yields the $\psi$-function of the second subsystem. The \underline{kind} of measurement determines the kind of $\psi$-function of the second subsystem. Different kinds of measurements on the first subsystem yield \underline{different kinds} of $\psi$-functions for the second subsystem.

3) Different $\psi$-functions describe different real states. (Prerequisite: the $\psi$-function describes a system \underline{completely}.) Therefore, the real state of the second subsystem depends on what kind of measurement I make on the first subsystem. This is equivalent to the assumption of an immediate long-distance-action. This was to be shown.

4) It is therefore not an internal contradiction of quantum theory, but the assertion that the following alternative exists

Either the description by quantum theory has to be considered as an incomplete description (of the individual system)

or one has to assume an immediate action-at-distance.

(Or one may also assume both).

Sincerely yours,\\
Albert Einstein.

}
\section{Comments on the correspondence}\label{sec_comments}

\subsection{Comments from a physicist's point of view}\label{sec_com_phys}

In his first letter written to Einstein, Cooper suggests that the EPR paradox can be removed by assuming that the two quantum systems, commonly called $A$ and $B$ (here I and II), are separated by an infinite barrier so that each system is restricted to a halfspace. However, the assumption that related observables like momentum, position and energy are then restricted to a halfspace is definitely wrong. For it is not compatible with the fundamental axioms of quantum theory. In the present situation it is the \emph{wave function} that is restricted to a halfspace. It is generally accepted that observables must be self-adjoint so as to give rise to a spectral resolution. Defined on a halfspace they are merely Hermitian operators without any physical interpretation whatsoever. Think of the suggested barrier as a very narrow but tall potential wall. Such a potential enters the Hamiltonian but has no influence on the two momenta as operators, though it has an influence on the momenta measured by an observer since the observed values are dictated by the two wave functions for A and B (almost) located in halfspaces provided the two systems are sufficiently far apart from each other.

Needless to say that the time evolution operator $U_t = e^{-itH}$ is unitary and defined for all real $t$, no matter what kind of potential barrier tries to separate $A$ and $B$. Therefore, any normal quantum physicist will reject the suggestion that in certain situations $U_t$ can only be defined for $t \ge 0$ or $t \le 0$ as claimed by Cooper. Though in statistical quantum mechanics one introduces the operator $e^{-\beta H}$ with bounded-below $H$, this operator makes sense only for $\beta \ge  0$.

It is interesting to see that Einstein in response to the letter by Cooper does not point at the contradictions and inconsistencies mentioned above. The reasons are twofold.

\medskip\noindent
(1) Einstein never got interested in the mathematical foundation of quantum mechanics. For him there was no reason to object or discuss Cooper’s mathematical proposals.

\medskip\noindent
(2) His main concern was the possible \emph{incompleteness} of quantum theory, not the \emph{logical inconsistency} of it. He accepted the current theory and all the results though they did not make him happy.\medskip

Cooper’s impassable barrier offers no solution to the EPR paradox. Of course, one has to avoid a considerable interaction between $A$ and $B$. But this can be achieved simply by separating these two systems by a large distance so that one avoids interaction. A potential barrier, no matter how large, is not  necessary to reduce the possibility of an interaction. For the conclusion that there will be in general \emph{no independent existence of spatially separated parts} it is helpful not to insist on a the presence of a potential barrier. Therefore, in his letter Einstein states that he does not want to suppose that there is a wall. He rejects Cooper’s suggestion.

Einstein formulates two contradicting principles:

\medskip\noindent
(a) Independent existence of spatially separated parts.\medskip

\medskip\noindent
(b) The $\psi$-function is the complete description of an individual physical situation.\medskip

His thesis that both cannot be true together is certainly correct. But his decision was that (a) is correct and (b) is not. Niels Bohr, shortly after 1935, claimed the opposite: (b) is correct while (a) is not. In order to understand Einstein’s viewpoint one has to look at the EPR paper published in \emph{Physical Review} 1935 \cite{EPR}. In the beginning the authors formulate:

\emph{Any serious consideration of a physical theory must take into account the distinction between the objective reality, which is independent of any theory, and the physical concepts with which the theory operates.}

This shows that Einstein believed in an objective reality behind the phenomenon, the observation, the description. It was precisely this viewpoint which Niels Bohr rejected. The term objective reality was meaningless to him and remains so until today for the majority of physicists. Indeed, the paradox in the EPR paper relies on the assumption of the existence of a reality behind the wave function. This reality, if it exists, needs a separate description providing further details of the object. As for today most physicists agree with Bohr’s opinion and consider the entire dispute as philosophy (see \cite{Jammer_1974}) or as a chapter of logic (see \cite[Chapter~9]{Popper_1959}. Already in 1927 Bohr emphasized that we have to treat with extreme care our use of language in recording the results of observations that involve quantum effects. Bohr’s deep concern with the role of language never ceased.

Experimentally the entanglement of noninteracting quantum systems has been confirmed and has become a standard subject of research \cite{Audretsch_2007}.

In response to Einstein’s letter Cooper wrote his second letter where he tried to explain why it seems necessary to him to separate the two particles
(\ie electrons) by an infinite barrier. He argues as follows. If each particle is known to be confined each to a finite interval of the real line, its wave function $\psi(x)$ vanishes outside that interval, so that its Fourier transform is analytic and so can have only isolated zeroes. Hence the momentum $p$ will have finite values in any interval, and the particles will interact, no matter how far apart they are. It is our careless use of language that caused this conflict. A particle is said to be \emph{localized} if its wave function decreases exponentially outside some small area of space. It seems impossible to create experimentally a situation where the wave function vanishes outside that area. The exponential decrease holds for instance also for an electron bounded within the hydrogen atom.

Cooper claims that in the EPR situation without a barrier an uncertainty arises as to whether particle $B$ has been in collision with particle $A$. It follows from his argument that it is the uncertainty about $A$ and $B$ moving together, and not any kind of force between the two, which causes the uncertain values for $x$ and $p$ of $B$ in the EPR \emph{Gedankenexperiment}. Only a barrier will make it impossible for $A$ and $B$ to move together.

One month later Einstein wrote back in German claiming not to have the time to write in English. As before, he refuses to accept the existence of any barrier, saying that it is of no interest for the questions he posed in the EPR paper and other papers published between 1935 and 1949. He repeats his conviction that quantum theory does not suffer from an internal contradiction. He repeats the discrepancy between the demand that there be no entanglement at large distances and the principle that the wave function gives a complete description of a quantum object. He also states that wave functions never vanish outside a bounded region, and so Cooper’s argument fails. He still uses the term \emph{real state} (realer Zustand) which suggests the existence of an \emph{objective reality} which we now consider doubtful.

\subsection{Comments from a mathematician's point of view}\label{sec_com_math}

The conflict between Einstein and Cooper is based on the different demands that natural scientists or engineers on the one hand and mathematicians on the other place on a mathematical model.

Let us first demonstrate this by means of a simple example from signal processing. Communication engineers mainly use the model of band-limited signals, i.\,e. the Fourier transform of the signal function $f$ disappears outside a finite interval. This means, however, that the function $f$ is analytic and thus has only isolated zeros, unless it vanishes identically. In particular, the signal cannot be time-limited. So, it has no beginning and no end, which contradicts the idea of a real signal.

When talking to a communication engineer about this contradiction, he will respond in a similar way as Einstein: ``Of course, a real signal cannot be precisely band-limited, but it can be thought to be realized with any precision and that is enough for practical applications.'' In fact, digital signal processing, based on the model of band-limited signals, is known to work very well.

Mathematicians, of course, cannot live with this contradiction. For this reason, they have studied signal processing for non-band-limited signals and, inter alia, investigated the error caused by replacing a non-band-limited signal with a band-limited one; see, \eg, \cite[Section~3.4]{Butzer-Splettstoesser-Stens_1988}, \cite{Beaty_et_al_2009,Brown_1967,Butzer-Stens_1992_SIAM,Weiss_1963} and the literature cited therein.

In his first letter of October~11, 1949, Cooper requires an impassable barrier between the two particles I and II, in order to avoid the EPR paradox. Without this barrier, there is a finite probability that I will move to II and disturb it (by interference) within any time interval, no matter how short, after a measurement on I is made. In his letter of November~19, Cooper states more precisely: ``Suppose that at an initial instant it is known definitely that two particles are on intervals of the $x$-axis which are as far apart as you like. Then, \dots, the wave functions in terms of $x$ vanish for each particle outside a finite interval, and the momentum wave function, being the Fourier transform of the $x$-wave function, is an analytic function, and so can have only isolated zeroes. There is therefore a finite probability that the momentum may have values in any interval, however large; consequently, there is a finite probability that the particles will be together, and hence interacting, at any instant after the initial instant, \dots'' See also \cite{Cooper_1950b}. 

In his reply of November~11, to Cooper’s second letter, Einstein answers in a similar way as engineers do in case of band limitation. The translation of his main argument written in German is: ``The analytical character of the $\psi$-function does not, of course, allow $\psi$ to disappear exactly for any finite $(x_1, x_2)$ domain.  But this can be thought to be realized with any approximation; This is sufficient for our reasoning. This means that we know for sure that the two subsystems are spatially separated.''

This is the same discrepancy as that between time and band limitation, now with the role of the function and its Fourier transform interchanged. Cooper certainly would not have accepted that a real, thus time-limited signal, can be band-limited at the same time. However, communication engineers, as seen above can live very well with this contradictory model. Similarly, Einstein knows that the $\psi$-function cannot vanish on an interval, but since it can ``approximately vanish'', he concludes that the two subsystems are spatially separated.

The problem that physicist have with the model of an impassible barrier lies in the fact that the operators representing the momenta are no longer self-adjoint, and hence the expansion involving their eigenfunctions is not valid. The eigenfunction expansion is an essential argument of the EPR paper. In Section~\ref{sec_com_phys}, the physicist of us writes: ``It is generally accepted that observables must be self-adjoint so as to give rise to a spectral resolution. Defined on a halfspace they are merely Hermitian operators without any physical interpretation whatsoever.''

The discrepancy between physical models and the exact mathematical descriptions has been of interest to Cooper throughout his whole life. I (RS) remember a lecture in Aachen in the seventies, based on his article “The Foundations of Thermodynamics” \cite{Cooper_1967}. He talked about the different forms of the second law of thermodynamics, which are  ``equivalent'' in the sense of physics. Indeed, physicists prove this equivalence by means of a so-called Carnot cycle (thermodynamic cycle), in which an ideal gas passes through different states and is at the end again in the initial state. On the other hand, physicists know that such an ideal gas does not exist, indeed cannot even exist, because such a gas would be contradictory in many respects. Cooper, as a mathematician, therefore, did not accept these proofs of equivalence. In \cite{Cooper_1967} he states ``\dots it is in any case illogical to base a physical theory on the assumption of existence of substances which do not exist in nature.” This view is also shared by many other mathematicians and mathematical physicists; see the literature cited in \cite{Cooper_1967}.

Again, for most physicists this discrepancy can be neglected. They argue analogously to Einstein: ``Of course, an ideal gas cannot exist, but it can be thought of with any approximation; this is sufficient for our considerations.''

%
%
%
%
%
%
%
%
%
%

\bibliographystyle{my_abbrvnat}
\bibliography{Cooper}


\end{document}